\documentclass[twocolumn,aps,prb,floatfix,showpacs,superscriptaddress]{revtex4}
\usepackage{graphicx,subfigure}
\usepackage{dcolumn}
\usepackage{bm}
\usepackage{psfrag}
\usepackage{amssymb,amsmath}

\begin{document}
  \title{Helical edge states in multiple topological mass domains}
  
\author{P. Michetti}
\affiliation{Institute of Theoretical Physics and Astrophysics, University of W\"urzburg, D-97074 W\"urzburg, Germany}

\email{michetti@physik.uni-wuerzburg.de, precher@physik.uni-wuerzburg.de}

\author{P. H. Penteado}
\affiliation{Instituto de F\'isica de S\~ao Carlos, Universidade de S\~ao Paulo, 13560-970, S\~ao Carlos, SP, Brazil}
\affiliation{Institute of Theoretical Physics and Astrophysics, University of W\"urzburg, D-97074 W\"urzburg, Germany}

\author{J. C. Egues}
\affiliation{Instituto de F\'isica de S\~ao Carlos, Universidade de S\~ao Paulo, 13560-970, S\~ao Carlos, SP, Brazil}

\author{P. Recher}
\affiliation{Institute for Mathematical Physics, TU Braunschweig, 38106 Braunschweig, Germany}
\affiliation{Institute of Theoretical Physics and Astrophysics, University of W\"urzburg, D-97074 W\"urzburg, Germany}

 \begin{abstract}
The two-dimensional topological insulating phase has been experimentally discovered in HgTe quantum wells (QWs).
The low-energy physics of two-dimensional topological insulators (TIs) is described by the Bernevig-Hughes-Zhang (BHZ) model, where the 
realization of a topological or a normal insulating phase depends on the Dirac mass being negative or positive, respectively.
We solve the BHZ model for a mass domain configuration, analyzing the effects on the edge modes of a finite Dirac mass in the normal insulating region (soft-wall boundary condition). 
We show that at a boundary between a TI and a normal insulator (NI), the Dirac point of the edge states appearing at the interface  
strongly depends on the ratio between the Dirac masses in the two regions.
We also consider the case of multiple boundaries such as NI/TI/NI, TI/NI/TI and NI/TI/NI/TI.
%
\end{abstract}

  \maketitle

  \section{Introduction}
  Topological insulators (TIs) are time-reversal-symmetric materials featuring a topological phase characterized by a $\mathbb Z_2$ topological invariant~\cite{kane2005a,Moore2007}. 
  In two-dimensions (2D), they exhibit the quantum spin Hall (QSH) phase \cite{kane2005a,kane2005b}.
  The QSH phase has been theoretically predicted~\cite{bernevig2006} and experimentally realized in HgTe/CdTe QWs~\cite{konig2007}. 
  The crucial ingredient of this narrow gap semiconductor material is the inverted band structure of HgTe.
  Similarly, 3D TIs supporting chiral fermions as surface states have been proposed 
  and observed~\cite{fuinversion2007,hsieh2008,zhang2009,hsieh2009,xia2009,chen2009}.

  In HgTe/CdTe QWs, the topological phase is determined by the sign of the Dirac mass $M$. 
  The gap between the E1 (s-like) and the H1 (p-like) subbands at the $\Gamma$ point is given by $2|M|$.
  The only experimentally accessible parameter tuning the Dirac mass from normal ($M>0$) to 
  inverted ($M<0$) is the thickness of the HgTe QW.
  In particular, a topological transition from the normal to the topological insulating phase takes place when the QW thickness is increased above the critical thickness $t_C=6.3$~nm~\cite{konig2007}.  
  Recently, electrically  driven topological insulating phase transitions have been proposed in heterostructures with gate tunable conduction-valence band energy separation.
  %
  In particular, in Ref.~\onlinecite{liu2008} a type-II InAs/GaSb/AlSb QW was proposed and recent experiments~\cite{knez,knez2} provided 
  the first evidence pointing towards the presence of a topological insulating phase in these structures. 
  In Ref.~\onlinecite{michetti2012}, double QW structures composed of narrow gap semiconductors are considered featuring a tunable topological transition with the application of a gate bias of the order of the gap of the individual
  trivial QWs.    
  Both proposals pave the way to 2D systems where mass domains are designed with lithographic gates to create topological and normal regions.
  For example, ring-shaped TI regions are particularly interesting for the peculiar properties of their confined edge states, 
  which can be controlled with a threading magnetic flux~\cite{Maciejko2010, Michetti2011}. 
  On the other hand, the Dirac mass term in single HgTe QWs is related to the QW thickness and therefore thickness fluctuations of less than $1$~nm height can 
  accidentally determine the formation of mass domains alternating TI regions, where $M<0$, to NI regions, where $M>0$.\cite{footnote1} 
  These phenomena can be especially relevant for near zero-gap HgTe QWs~\cite{Buttner2011}. 
\begin{figure}[tbh]  
    \includegraphics[width=7.5cm]{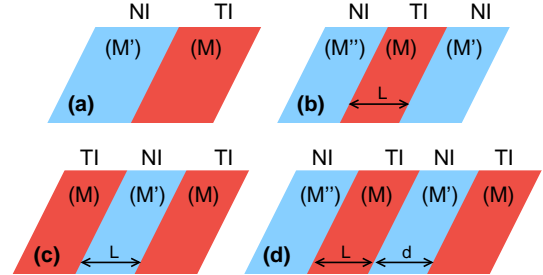}
  \caption{(Color online)
    Schemes of multiple topological mass domains analyzed in the present paper.
    According to the classification of the topological phase of each region into a TI or a NI we have $M<0$ or $M'>0$ and $M''>0$, respectively.
}
  \label{fig:domains}
\end{figure}

  The change in the topological invariant $\mathbb Z_2$ between two systems determines the presence of 1D helical edge states running along the boundary between the TI and the NI regions, 
  a phenomenon referred to as the bulk-boundary correspondence. 
  %
  Such edge states are topologically protected against single particle elastic backscattering (as long as time reversal symmetry [TRS] is preserved) 
  and are particularly interesting for their spin and charge transport properties.  
  In the literature, these edge states are generally obtained by solving the Hamiltonian with \emph{hard}-wall boundary conditions (BCs), i.e. by  
  imposing that the wave function \emph{vanishes} at the interface between a TI and a normal medium (an exception is the recent proposal of \emph{natural} boundary conditions~\cite{Medhi2012}).
  While this condition is appropriate to treat the interface between TIs and the  $M\rightarrow+\infty$ vacuum, 
  this is by no means a good approximation when dealing with electrically-induced mass domains~\cite{liu2008,michetti2012}, 
  where the normal regions have a finite positive Dirac mass. 
  Finite mass-domains in Dirac systems have first been solved in the context of zero energy bound 
  states in the 1+1 Dirac equation by Jackiw and Rebbi~\cite{Jackiw76} and for interface states in band-inverting contacts based on HgCdTe and PbSnTe ~\cite{Volkov85, Pankratov87}.
  More recently, finite mass-domains were proposed to induce valley-polarized metallic states in biased bilayer graphene~\cite{Martin2008}.

  In the present paper we solve the edge states of a HgTe/CdTe QW-based TI for the case of \emph{soft}-wall BCs, 
  appropriate for describing systems with Dirac mass domains, where the wave function does \emph{not} vanish at the interface, 
  but its continuity and the continuity of its normal derivative are instead required.
  In Section II, we briefly review the BHZ model and describe the method used to solve the edge states in the topological mass domains.
  In Section III, we deal with a single NI/TI interface schematized in Fig.~\ref{fig:domains}(a), with both hard-wall and soft-wall BCs.
  We show that soft-wall BCs  quantitatively change the dispersion curves [Fig.2(a)] with respect to the case of hard-wall BCs.
  In particular, the hard-wall limit is only reproduced in the limit $M'\rightarrow\infty$.
  Even for $M'=10^4$~meV, which is of the order of magnitude of the electron extraction work-function of a crystal, an appreciable deviation from the hard-wall limit is still observed.
  The density profile of the edge state bound to the mass-domains (Fig.~3) is qualitatively affected by the soft-wall BCs, which allow it to extend into both the TI and the NI regions.
  In Section IV, we address the effect of a finite bulk inversion asymmetry (BIA) term, which introduces off-diagonal matrix elements to the otherwise block diagonal form of the BHZ model, and 
  evaluate its effects on the bulk dispersion curves and on the edge state dispersion (Fig.~\ref{fig:BIA}). 
  The effect of BIA on the edge states of a single NI/TI is shown to be tiny, so that BIA can be safely neglected in the more complex case of multiple NI/TI boundaries. 
  In Section V, we analyze a system with two TI/NI interfaces. 
  In particular, we investigate the case of a TI strip embedded in a normal system, sketched in Fig.~\ref{fig:domains}(b), and vice versa: the case of a TI system where a strip region with $M>0$ is 
  present [Fig.~\ref{fig:domains}(c)] and analyze for the first time the edge coupling through the normal region.
  In both cases, the overlap of edge states bound to different boundaries leads to a fully gapped edge mode spectrum (Fig.~\ref{fig:sw1} and Fig.~\ref{fig:lateral}), 
  with a minigap exponentially shrinking with the distance between the two interfaces. 

  Soft-wall BCs can also be used to couple three or more edge states, thus, in Section VI, we analyze 
  the edge states for a system with three NI/TI boundaries [Fig.~\ref{fig:domains}(d)].  
  We study how the properties of this system vary with the Dirac mass and geometrical parameters.
  We further argue that this system can describe a helical edge state at the sample boundary (vacuum/TI interface) in the presence of Dirac mass 
  fluctuations in the TI composition giving rise to mass domains with bubbles having $M>0$ (i.e. normal character) in the bulk of 
  the sample whose edge states could interact with the helical edge states at the sample boundary.  

  \section{The BHZ model}
  The spectrum of a HgTe QW near the $\Gamma$-point is effectively described in its low energy sector~\cite{schmidt2009} 
  by the 4-band model~\cite{bernevig2006} 
  \begin{eqnarray}
    H_{\vec k}&=& \left(\begin{array}{cc}
      h({\vec{k}}) &0\\
      0 & h^*(-\vec{k})
      \end{array}\right)\nonumber\\
    h(\vec k)&=&  \vec{d}\cdot\vec{\sigma}\label{eq:H0}\\
    \vec{d}&=& \left( \varepsilon_k, A k_x , -A k_y   , M_k \right)\nonumber\\
    \varepsilon_k &=&C-D k^2 \hspace{1cm}   M_k= M-B k^2, \nonumber
  \end{eqnarray}
  where $k=|\vec k|=\sqrt{k_x^2+k_y^2}$ and $\vec{\sigma}$ is the vector of Pauli matrices associated with the band-pseudospin degree of freedom (band $E_1$ or $H_1$) \cite{footnote2}.
  $H_{\vec{k}}$ is represented in the basis $\big\{|E_1+\rangle$, $|H_1+\rangle$, $|E_1-\rangle$, $|H_1-\rangle\big\}$, 
  where the $E_1$  states ($J_z=\pm1/2$) are a mixture of the s-like $\Gamma_6$ band with the $\Gamma_8$ light-hole 
  band, while $H_1$ ($J_z=\pm 3/2$) is basically the $\Gamma_8$ heavy-hole band. 
  For later use in numerical simulations, we quote the following choice of parameters: $A=375$ meV nm, $B=-1120$ meV ${\rm nm}^2$ 
  and $D=-730$ meV ${\rm nm}^2$. These parameters follow from the 8x8 Kane model~\cite{novik2005}.
Without loss of generality we also assume $C=0$.
 %
  %
  The Dirac mass $M$ depends on the QW thickness and $M<0$ corresponds to the inverted (QSH) regime whereas  $M>0$ corresponds to the normal regime.
  In a first approximation, $H_{\vec k}$ is block diagonal in the spin degree of freedom~\cite{bernevig2006}, for which we define the corresponding vector of Pauli matrices $\vec{\tau}$.
  As we consider only systems with TRS, we can restrict ourselves to the block $h({\vec{k}})$.
  Results can be extended to the other Kramers block $h^{*}(-\vec{k})$ which is related to $h(\vec{k})$ by the time reversal operation $\hat{T}=i \tau_y \sigma_0 \hat{K}$, where $\hat{K}$ is the operator of complex conjugation.

  The bulk dispersion curves obtained as the eigenvalues of Eq.~(\ref{eq:H0}) are described by 
  \begin{equation}
    E_\pm(\vec k) = \varepsilon_k \pm \sqrt{(M - B k^2)^2 + A^2k^2}.
    \label{eq:bulk}
  \end{equation}
  With a standard choice for the TI parameters (like the parameters stated above) the bulk dispersion relation displays a conduction band minimum (valence band maximum) 
  at $k=0$ with energy $E=M$ (or $E=-M$).
  However, depending on the values of the parameters in Eq.~(\ref{eq:H0}), the bulk dispersion curves can also show a ``Mexican hat'' behavior.
  For a detailed analysis of the behavior of the 4-band model as a function of its parameters, see Appendix~\ref{app1} and Ref.~\onlinecite{Lu2012}.

  \subsection{Boundary conditions \label{procedure}}
  We are interested in obtaining the eigenstates of Eq.~(\ref{eq:H0}) in real space for a semi-infinite geometry, invariant under translations along the $x$-axis.
  For such a system, $k_y$ is no longer a conserved quantity and should be replaced by the operator $-i\partial_y$.
  Compatible with a fixed energy $E$ and a real $k_x$, the secular equation $|h({\vec k})-E|=0$ provides four $k_y$-modes:
  \begin{equation}
    k_y^2 \equiv k_\pm^2 = -k_x^2 -F \pm \sqrt{F^2-Q^2}
    \label{modi}
  \end{equation}
  with
  \begin{eqnarray}
    F &=& \frac{A^2 -2(B M+DE)}{2(B^2-D^2)},\\
    Q^2 &=& \frac{M^2-E^2}{B^2-D^2}.
  \end{eqnarray}
  With our choice of parameters, $k_\pm$ has one imaginary and one real solution for $E$ within the energy range of the bulk bands, whereas in the bandgap $\left(-|M|,|M| \right)$ 
  both values for $k_y$ are imaginary. 

  For each $k_y$-mode, one can write the spinors satisfying the Schr\"odinger equation and corresponding to the Kramers blocks $\tau=\pm 1$ as 
  \begin{equation}
    \psi_{k_x, k_y,\tau}(x,y) = \frac{e^{i k_x x}}{\sqrt{L_x}} 
    \left(\begin{array}{l}
        e^{i k_y y}\\
        R_{\tau, k_y} e^{ i k_y y}
      \end{array}
    \right).
    \label{eq:spinor0}
  \end{equation} 
  The ratio between the two components is
  \begin{equation}
    R_{ \tau, k_y} = - \frac{A (\tau k_x - i k_y)}{-M - E - (D-B) (k_x^2+k_y^2)}.
    \label{eq:erre}
  \end{equation}
  The general solution of the Dirac equation with energy $E$ and wave vector $k_x$ is therefore given by a linear combination 
  of the four solutions $k_y=\lambda k_\mu$ (with $\lambda,\mu=\pm$) obtained from Eq.~(\ref{modi}), 
  \begin{equation}
    \Psi_{k_x, \tau}^{(n)}(x,y) =  
    \sum_{\lambda, \mu=\pm}
    c_{\lambda,\mu}^{(n)}
    \psi_{k_x,\lambda k_\mu, \tau}(x,y),
    \label{eq:spinor}
  \end{equation}
  where we have introduced the index $n$ to refer, in what follows, to the $n$-th mass domain region.

  Here we discuss the general procedure we use to solve the BHZ model in a system composed by $N$ mass domains with parallel boundaries at $y=y_n$ with $n=0, 1, \dots N$.
  Inside the $n$-th mass domain with the condition $y \in \left(y_{n-1},y_n\right)$, we consider the Dirac mass term (and all other parameters) 
  as constant and a general expression for the spinor is given by Eq.~(\ref{eq:spinor}).
  At $y=y_n$ the value of the Dirac mass $M$ changes step-like~\cite{adiabatic}.
  
  Hard-wall BCs at $y$ for the domain $n$ are expressed by 
  \begin{equation}
    \Psi_{k_x, \tau}^{(n)}(x,y)=0 \hspace{0.5cm} \forall x,\label{hard}  
  \end{equation} 
  meaning that the edge state cannot extend beyond the boundary, being subject to a hard-wall confinement.  
  Soft-wall BCs between two consecutive domains $n-1$ and $n$ are instead expressed by the continuity of the 
  spinor and its normal derivative 
  \begin{eqnarray}
    \Psi_{k_x, \tau}^{(n-1)}(x,y_n)= \Psi_{k_x, \tau}^{(n)}(x,y_n)\hspace{0.5cm} \forall x,\nonumber\\
    \partial_y \Psi_{k_x, \tau}^{(n-1)}(x,y_n)= \partial_y \Psi_{k_x, \tau}^{(n)}(x,y_n)\hspace{0.5cm} \forall x. \label{soft}
  \end{eqnarray} 
 
  A system of BCs (either soft or hard) can be always expressed in a compact form as 
  \begin{equation}
    \mathbb{M}_{k_x}(E)~\vec{c}=0,
    \label{boundary_c}  
  \end{equation} 
  where $\vec{c}$ is a vector containing all the free coefficients $c_{\lambda,\mu}^{(n)}$ characterizing the wave function [Eq.~(\ref{eq:spinor})] in the $n=1,\dots N$ mass domain.
  $\mathbb{M}_{k_x}(E)$ can be constructed by appropriately using either Eq.~(\ref{hard}) or Eq.~(\ref{soft}) at each one of the boundaries of the system.
  As in standard quantum mechanics, the BCs determine the eigenenergies $E_{k_x}$ through the secular equation
  \begin{equation}
    \det{\left[\mathbb{M}_{k_x}(E)\right]}=0. 
    \label{det}
  \end{equation}
  The corresponding eigenspinors we obtain by solving Eq.~(\ref{boundary_c}) for the coefficients $\vec{c}$. 

\begin{figure}[tbh]  
  \subfigure{
    \includegraphics[width=6.5cm]{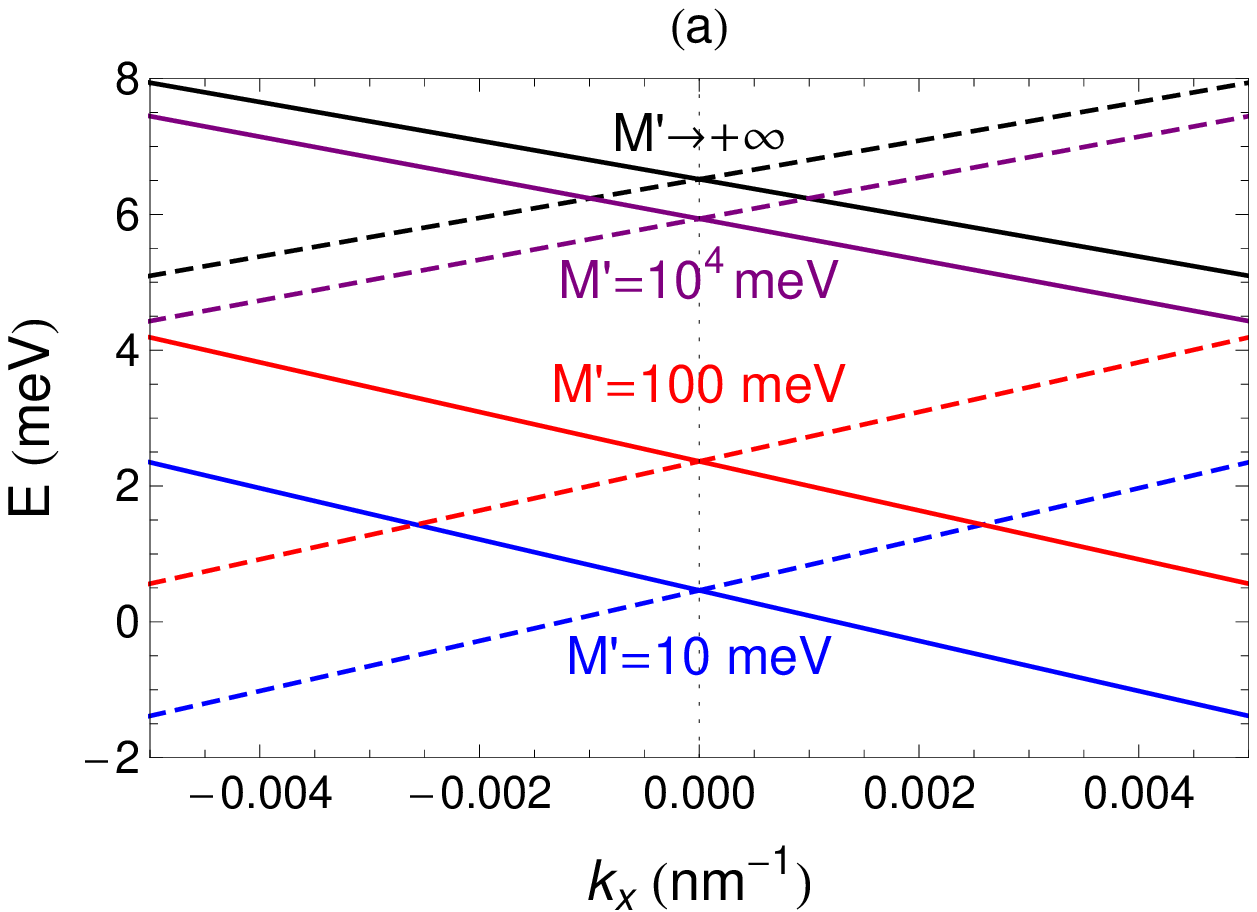}
    \label{fig:energy}
  }\vspace{0.4cm}
  \subfigure{
    \includegraphics[width=6.8cm]{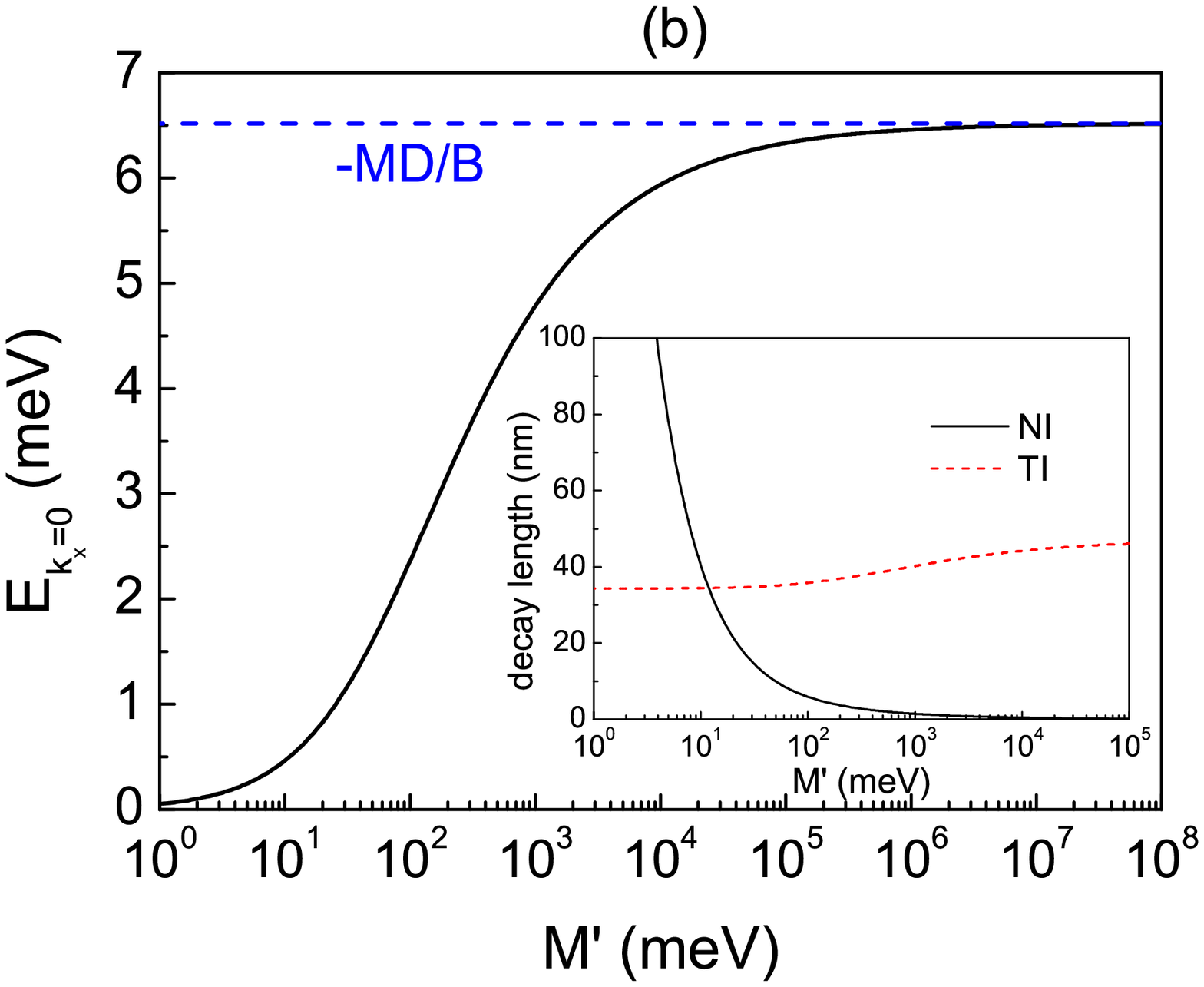}    
    \label{fig:Dirac}
  }
  \caption{(Color online)
    (a) Energy dispersion relation of helical edge states at the interface between 
    a TI region with Dirac mass $M=-10$~meV and a NI region of mass $M'$. 
    Solid lines and dashed lines correspond to the two Kramers blocks of the Hamiltonian Eq.~(1). 
    (b) Value of the Dirac point $E_{k_x=0}$ as a function of $M'$ (solid line). 
    The dashed red line shows the hard-wall limit $M'\rightarrow +\infty$, $E_{k_x=0}=-MD/B$.
    In the inset, edge states' characteristic decay length [$|k_+|^{-1}$ in Eq.~(\ref{modi})] in the 
    NI (full curve) and in the TI (dashed curve) plotted as a function of $M'$ at $k_x=0$ are shown.   
}
  \label{fig:oneedge}
\end{figure}

  \section{An isolated boundary}
  %
  %
  In this section, we consider a single interface between a TI ($y>0$) with $M<0$ and a NI ($y<0$) with $M'>0$ [see Fig.~\ref{fig:domains}(a)] and calculate the resulting helical edge modes.
  In the limit $M'\rightarrow +\infty$, we recover the usual hard-wall BCs of Eq.(\ref{hard}).
  For a finite positive $M'$, soft-wall BCs of Eqs.~(\ref{soft}) are employed.

  \subsection{Hard-wall boundary conditions}
  We apply vanishing BCs at $y=0$ and search for modes with energy lying within the bandgap.
  Only two of the four $k_y$ solutions of Eq.~(\ref{modi}) with positive imaginary part, which we define as $\tilde{k}_\pm$, 
  are normalizable in the region $y>0$ and contribute to the edge states.
  When $k_\pm$ are purely imaginary, we define $\tilde{k}_\pm = k_\pm$.
  When the parameters are such that $k_\pm$ are complex (see in Appendix~\ref{app1} Eq.~\ref{complex}), we instead have $\tilde{k}_\pm = \pm k_\pm$.
  Using Eqs.~(\ref{hard}) and (\ref{det}), we obtain the relation
  \begin{eqnarray} 
    R_{\tau,  \tilde{k}_+}&=&R_{\tau, \tilde{k}_-},
    \label{eq:come}
  \end{eqnarray}
  that imposes a strict relation between the energy $E$ and momentum $k_x$.
  Isolating $k_x$ terms and squaring twice, we arrive after some algebra at the edge mode dispersion curves~\cite{zhou2008, wada2011}
  \begin{eqnarray}
    E_{k_x} &=& -\frac{D}{B}M \pm k_x A\sqrt{\frac{B^2-D^2}{B^2}}.
    \label{solution}
  \end{eqnarray}
  First of all, we observe that after the first squaring of Eq.~(\ref{eq:come}), we lose track of $\tau$, 
  therefore only one of the $\pm$ signs in Eq.~(\ref{solution}) is actually a solution of Eq.~(\ref{eq:come}) for a given $\tau$.
  More important, because of the second squaring, the solutions in Eq.~(\ref{solution}) are not always allowed.
  For the usual parameter choice describing TIs, solutions are always admissible with $M<0$~\cite{zhou2008}. For $M>0$, the system admits no edge modes.

\begin{figure}[tbh]
  \centering	
  \includegraphics[width=7cm]{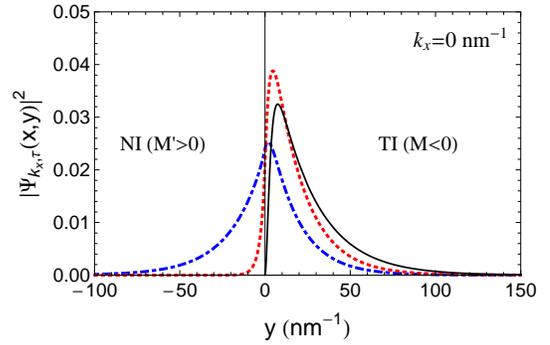}
  \caption{(Color online)
    Probability density of the edge state for $k_x=0$ as a function of the distance from the NI/TI interface ($y=0$).
    The different curves represent the hard-wall (solid line) and soft-wall (dashed lines) boundary condition cases with $M'=10$, $100$ meV for a fixed TI mass $M=-10$ meV. The probability densities  for the two blocks $\tau=\pm 1$  are degenerate.
  }
  \label{fig:onespinor}
\end{figure}

  \subsection{Soft-wall boundary conditions}
As mentioned previously, soft-wall BCs imply that at the interface ($y=0$) the spinor and its normal derivative are both continuous [see Eq.~(\ref{soft})]. 
For the $M'>0$ domain, i.e., $y<0$, only the $k_y$ modes of Eq.~(\ref{modi}) with negative imaginary part are allowed, while for $y>0$ possible solutions 
contain the modes with Im$\left(k_y\right)>0$. 
Using Eqs.~(\ref{soft}) and (\ref{boundary_c}) we obtain the $4\times4$ matrix
\begin{equation}
\mathbb{M}_{E,k_x}=\left(
\begin{array}{cccc}
1 & 1 & - 1 & - 1 \\
R_{\tau, -\tilde{k}_+} & R_{\tau, -\tilde{k}_-} & -R'_{\tau,\tilde{k}_+} & -R'_{\tau, \tilde{k}_-} \\
\tilde{k}_+ & \tilde{k}_- & \tilde{k'}_+ & \tilde{k'}_- \\
\tilde{k}_+ R_{\tau, -\tilde{k}_+} & \tilde{k}_- R_{\tau, -\tilde{k}_-} & \tilde{k}_+ R'_{\tau,\tilde{k}_+} & \tilde{k}_- R'_{\tau, \tilde{k}_-} \\
\end{array}
\right),
\label{eq:2edges}
\end{equation}
where the prime stands for both $\tilde{k}_{\pm}$ and $R_{\tau, \tilde{k}_{\pm}}$ calculated in the $M'$ domain. 
By numerically solving $\det{[\mathbb{M}_{E,k_x}]}=0$, we determine the energy dispersion relation of the helical edge states. 
In Fig.~\ref{fig:energy} we show the energy dispersions $E_{k_x}$ for a NI/TI interface keeping the TI Dirac mass $M=-10$ meV and varying the NI mass $M'$. 
Solid and dashed lines correspond to helical edge states of the spin-blocks $\tau$ and $-\tau$, respectively. 
The slope of the curves (velocity $v_x$) is not altered by varying $M'$. 
However, as shown in  Fig.~\ref{fig:Dirac}, the Dirac point rises with increasing $M'$ from $E_{k_x=0}=0$ (its limiting value for $M'\rightarrow0$) and 
eventually saturates at the value $E_{k_x=0}=-M\frac{D}{B}$ [see Eq.~(\ref{solution})] for $M'\rightarrow\infty$, reproducing the hard-wall case. In the inset of Fig.~\ref{fig:Dirac}, we plot the decay length in the TI and NI regions as a function of $M'$.

Figure~\ref{fig:onespinor} displays the corresponding probability densities $|\Psi_{k_x,\tau}\left(x,y\right)|^2$ of edge states at the Dirac point 
for soft-wall BCs with $M'=10$, $100$~meV (dashed lines) and hard-wall BCs (solid line). 
In both cases, the wave functions are strongly peaked closely to the interface $y=0$ and exponentially decaying away from the interface. 
The characteristic decay length of the edge states are given by the inverse of the smaller $k_y$ mode calculated from Eq.~(\ref{modi}) compatible with their eigenenergies and $k_x$ values.

\section{BIA effects \label{BIA}}
In the present section we address how bulk inversion asymmetry (BIA) affects the TI Hamiltonian in Eq.~(\ref{eq:H0}) and the edge mode solutions for hard-wall boundary conditions using the model of Ref.~\onlinecite{konig2008}.
We show that such BIA introduces a weak non-linearity in the edge dispersion, especially near the bulk band edges.
The position of the Dirac point is not affected while the velocity of the helical particle is only slightly modified. We therefore will not consider the effect of BIA in the sections that follow. 

\begin{figure}[tbh]
  \centering	
  \includegraphics[width=6.5cm]{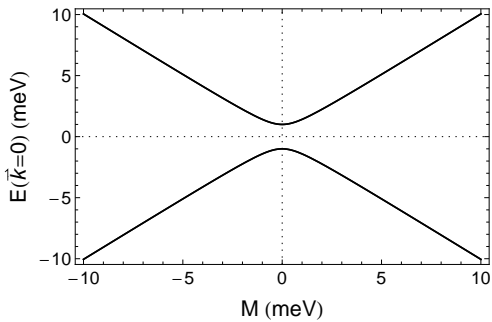}
  \caption{(Color online)
    Bulk-energy spectrum at ${\vec k}=0$ including the BIA-Hamiltonian Eq.~(\ref{eq:BIA}) as a function of Dirac mass parameter $M$. It shows an anticrossing with gap $2|\Delta|$.}
  \label{fig:anticrossing}
\end{figure}

\begin{figure}[tbh]
  \centering	
  \vspace{0.4cm}
  \includegraphics[width=6.8cm]{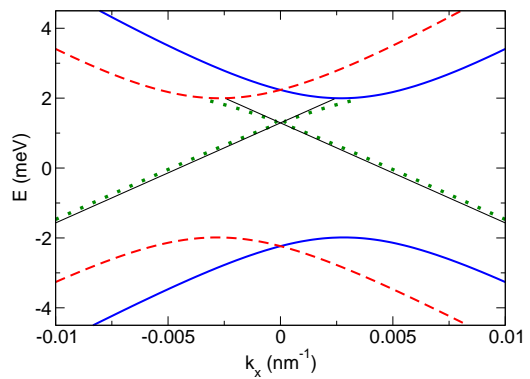}
  \caption{(Color online)
    Bulk bands dispersions including the effect of BIA terms, where full and dashed lines distinguish opposite helicities. 
    Edge states for a single hard-wall boundary, numerically obtained by solving Eq.~(\ref{eq:BIABHZ}) are shown in dotted lines. 
    The edge states dispersions for the corresponding system without BIA, obtained with Eq.~(\ref{solution}), are also displayed for comparison (full narrow line).
    We used the following parameters: $M=-2$~meV, $\Delta=1$~meV.
  }
  \label{fig:BIA}
\end{figure}

The block diagonal form of the BHZ model is indeed an approximation for the low-energy physics of a HgTe QW.
The presence of BIA introduces a coupling between $|E1,\pm\rangle$ and $|H1,\mp\rangle$ bands. 
The leading-order BIA perturbation term is expressed as~\cite{konig2008} 
\begin{equation}
  H_{BIA} = \left(
  \begin{array}{cccc}
    & & &-\Delta \\
    & &\Delta & \\      
    &\Delta&& \\
    -\Delta&&&
  \end{array}
  \right).
  \label{eq:BIA}
\end{equation}
It preserves TRS and therefore does not affect the topological properties of the BHZ model Eq.~(\ref{eq:H0})~\cite{konig2008}, 
as long as the bulk gap is not closed. Note that Eq.~(\ref{eq:BIA}) introduces an anticrossing at ${\vec k}=0$ as a function of $M$, 
see Fig.~\ref{fig:anticrossing}. Such an anticrossing has been also found in Ref.~\onlinecite{Winkler2012} as a function of the QW thickness and the inversion 
crossing of the $E_1$- and $H_1$-bands shifting to finite ${\vec k}$-values. Here we show (see Fig.~\ref{fig:BIA}) explicitly, that the helical edge states 
are still present in the model considered and only slightly modified, despite the anticrossing at ${\vec k}=0$.

Let us first rewrite the BHZ Hamiltonian in Eq.~(\ref{eq:H0}), including the BIA term in Eq.~(\ref{eq:BIA}), in the following form
\begin{equation}
  H =  \varepsilon_k I_{4\times 4} + A \vec{k} \vec{\Sigma} + \Delta \Lambda_y - M_k \Lambda_z \Sigma_z,
  \label{eq:BIABHZ}
\end{equation}
where we have introduced two sets of unitary and Hermitian matrices
\begin{eqnarray}
  \Sigma_x = \tau_z \sigma_x ;\hspace{0.5cm} \Sigma_y = -\tau_0 \sigma_y;\hspace{0.5cm} \Sigma_z = -\tau_z \sigma_z \nonumber\\
  \Lambda_x = \tau_x \sigma_y ;\hspace{0.5cm} \Lambda_y = \tau_y \sigma_y;\hspace{0.5cm} \Lambda_z =  \tau_z \sigma_0
  \label{eq:matrices}
\end{eqnarray}
with the property that each set separately obeys Pauli commutation rules, while elements from the two sets commute.  
Note that the only  matrices which are off-diagonal in the Kramers block pseudospin (i.e. containing $\tau_x$ or $\tau_y$) are $\Lambda_x$ and $\Lambda_y$.
We now perform the following unitary transformation in the $\Lambda$-$\Sigma$ space which warrants Eq.~(\ref{eq:BIABHZ}) block-diagonal: 
\begin{equation}
  U = \frac{1}{\sqrt{2}} \left[ -i(\zeta_y \Sigma_x -\zeta_x \Sigma_y ) \Lambda_y + \Sigma_z \Lambda_z \right]
  \label{eq:U}
\end{equation}
with $\vec{\zeta}=\vec{k}/|\vec{k}|$.
After the transformation the Hamiltonian in Eq.~(\ref{eq:BIABHZ}) acquires the following form
\begin{eqnarray}
  H_{\vec k} &=&  \varepsilon_k I_{4\times 4} +  \left(-A |\vec{k}| I_{4\times 4} +  \Delta \Lambda_z \right) \vec{\zeta} \vec{\Sigma} - M_k \Lambda_z \Sigma_z\nonumber\\
  &=&\left(
  \begin{array}{cc}
    h_{+}(k) &  0\\ 
    0 & h^*_{-}(-k)
  \end{array}
  \right),
  \label{eq:BHZ}
\end{eqnarray} 
where we have introduced the helicity parameter $\eta=\pm 1$, and have defined
\begin{eqnarray}
  h_{\eta}(k) &=&  \varepsilon_k I_{2\times 2} +  \left(-A |\vec{k}| +  \eta \Delta \right) \left(\zeta_x \sigma_x -\zeta_y \sigma_y \right) + M_k \sigma_z.\nonumber\\
\end{eqnarray} 
%
The helicity of the energy eigenstates is defined in the new basis by $\eta =\langle\tau_z\rangle$, which in the original basis of Eq.~(\ref{eq:BIABHZ}) 
is equivalent to $\eta =-\langle \left(\Sigma_x \zeta_x + \Sigma_y \zeta_y \right) \Lambda_y \rangle$.

%
The bulk dispersion curves obtained from Eq.~(\ref{eq:BHZ}) are shown in Fig.~\ref{fig:BIA}.
Similarly to an electronic system in the presence of the Rashba spin-orbit interaction, the dispersion curves can be classified through the helicity $\eta$. For a given wave vector, the effect of the BIA term is to lift the degeneracy of the two spin-blocks. 

In order to solve for the edge states of the system with BIA, we need to treat Eq.~(\ref{eq:BIABHZ}) in real space 
[we note that Eqs.~(\ref{eq:U}) and therefore Eq.~(\ref{eq:BHZ}) are well defined only in momentum space].
We follow the procedure illustrated in Section~\ref{procedure}, applying it to the $4\times4$ Hamiltonian in Eq.~(\ref{eq:BIABHZ}).
In Fig.~\ref{fig:BIA}, we show the edge states obtained for a system with $M=-2$~meV and $\Delta=1$~meV (dotted lines) and compare them with the edge states of 
the corresponding system with no BIA (full narrow lines), obtained analytically with Eq.~(\ref{solution}).
These BIA terms do not change the position of the Dirac point, but slightly change the group velocity close to the Dirac point. 
Away from the Dirac point the edge dispersion in Fig.~\ref{fig:BIA} shows a weak non-linear distortion, accentuated near the bulk band edges. Projecting $H_{BIA}$ 
onto the unperturbed edge states, it is straightforward to show that the effects of $H_{BIA}$ are at least of order $\Delta^2$, see Ref.~\onlinecite{Virtanen12}.

\begin{figure}[bph]
  \centering	
  \includegraphics[width=6.3cm]{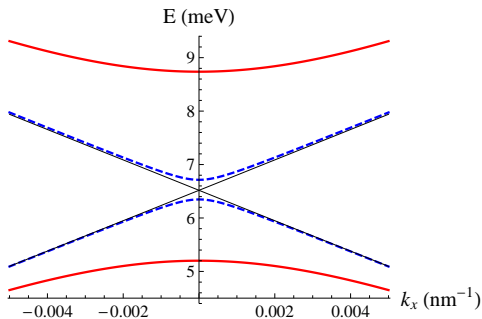}
  \caption{(Color online)
    Energy dispersion of a TI strip (NI/TI/NI) calculated with hard-wall boundary conditions.
    The solid red lines correspond to $L=100$~nm, blue dashed curves to $L=200$~nm and 
    black thin lines to $L=1000$~nm.
  }
  \label{fig:hw}
\end{figure}

\section{Systems with two boundaries}

\subsection{A strip of TI}
Here, we consider a NI/TI/NI mass domain shown in Fig.~\ref{fig:domains}(b), where Dirac masses are $M'>0$, $M<0$ and $M''>0$, respectively.
For very large $M'$ and $M''$ the use of hard-wall BCs is appropriate. For the entire subsection, the Dirac mass for the TI domain is $M=-10$ meV.
\subsubsection{Hard-wall boundary conditions}
The case of a TI strip confined by hard-wall BCs has been first analyzed by Zhou et al.~\cite{zhou2008}.
We briefly comment in this section some of their results in order to set a benchmark for successive extension to soft-wall BCs. 
The TI strip has two pairs of helical edge states (it is not topologically protected) exponentially localized at the two boundaries 
which are separated by the width of the TI strip $L$.
The decrease of the TI strip width leads to a finite overlap of edge modes belonging to different interfaces originating a minigap (a full gap 
in the edge mode dispersion curves) as shown in Fig.~\ref{fig:hw}. 
For $L=1000$~nm (thin full lines), the overlap is negligible, the minigap is exponentially suppressed.
The dispersion curves are linear, just two copies of single-interface edge modes shown in Fig.~\ref{fig:energy} for $M'\rightarrow\infty$.
For $L=200$ and $100$~nm, the overlap is instead substantial and the edge modes anticross at $k_x=0$ giving rise to a finite minigap.  
%
\begin{figure}[tbph]
    \centering	
    \includegraphics[width=5.5cm]{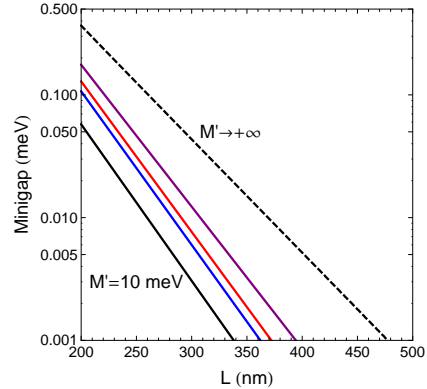}
    \caption{(Color online)
      Logarithmic plot of the minigap value as a function of the width $L$ of the TI strip for 
      soft-wall boundary conditions (solid lines) with $M'=M''=10$, $50$, $100$ and $350$~meV. The dashed curve corresponds to the hard-wall boundary condition.
    }
    \label{fig:sw1}
\end{figure}
%
\begin{figure}[tb]
    \centering	
   \vspace{0.2cm}
  \includegraphics[width=6.5cm]{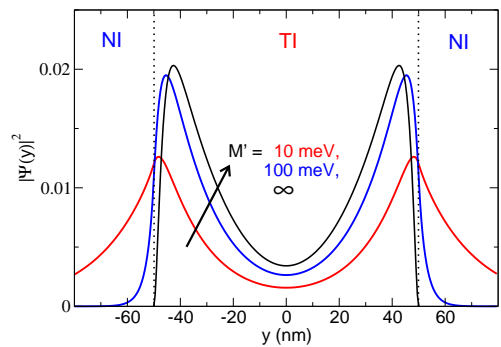}    
  \caption{(Color online) Electronic density of the edge mode of a TI strip of width $L=100$~nm at $k_x=0$. 
  Considered NI Dirac masses are $M'=M''=10$, $100$~meV and $\infty$ (hard-wall confinement).} 
   \label{fig:2edgesspinor}  
\end{figure}

\subsubsection{Symmetric soft-wall boundary conditions}
We now consider $M'=M''$ to be finite, adopt soft-wall BCs and solve for the edge modes.
The edge dispersion curves are qualitatively similar to the hard-wall case in Fig.~\ref{fig:hw}, 
and an anticrossing behavior  is found at $k_x=0$, with the opening of a minigap around the Dirac point of
the corresponding one-boundary edge modes.
We note that such an anticrossing point scales in energy  with $M'$, similarly to the behavior of the Dirac point with soft-wall BCs shown in Fig.~\ref{fig:Dirac}.
We plot in Fig.~\ref{fig:sw1} the minigap's exponential decay as a function of the TI strip width $L$. 
The dashed line corresponds to the hard-wall case, and the soft-wall cases with finite Dirac masses of $350$, $100$ , $50$ and $10$~meV are also shown.
With soft-wall BCs, the minigap is smaller and subject to have a faster decay, whose origin can be easily understood by looking at the edge states profile in Fig.~\ref{fig:2edgesspinor}. 
Here, we plot the profile of the electronic density of the $k_x=0$ edge modes in the direction perpendicular to the boundaries, for $L=100$~nm and for different NI masses 
ranging from $10$~meV to $\infty$ (hard-wall case).
As one increases $M'$ the edge modes become more strongly confined into the TI strip, 
thus leading to an enhancement of the overlap between the edge modes at the two interfaces (signaled by the increase of the probability at $y=0$).

\subsubsection{Asymmetric confinement}
\begin{figure}[tb]
    \centering	
    \includegraphics[width=6.8cm]{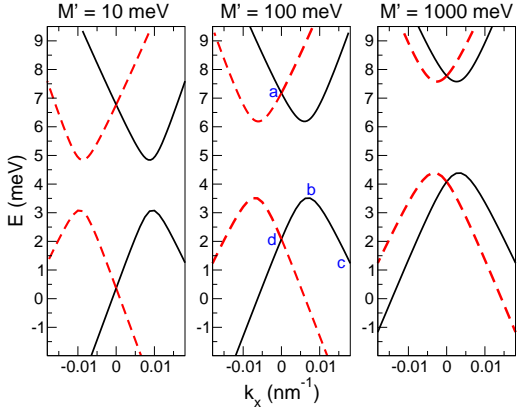}
    \caption{(Color online) 
       Edge modes of a TI strip with hybrid confinement:  $M''\rightarrow+\infty$ and $M'=10$, $100$, $1000$~meV for a TI region of width $L=100$~nm. The full and dashed lines distinguish $\tau=\pm 1$.
    }    
    \label{fig:hybridmodes}
\end{figure}

\begin{figure}[tbph]
    \centering	
    \vspace{0.3cm}
    \includegraphics[width=6.8cm]{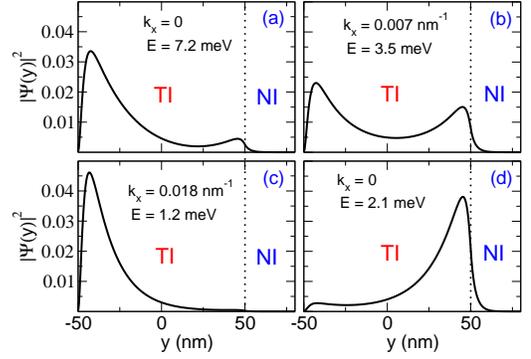}
    \caption{(Color online) Hybrid case -- one hard-wall ($M''\rightarrow+\infty$) and one soft-wall ($M'=100$ meV) with $L=100$ nm. The panels (a)-(d) correspond to states labeled by a-d in Fig.~ \ref{fig:hybridmodes}.}    
    \label{fig:hybridspinor}
\end{figure}

We consider now NI/TI/NI mass domains which differ from the previous ones for the lack of the mirror symmetry at the center of the TI strip.
This is the case whenever $M'\ne M''$.
In particular we will focus on the prototypical situation of a hybrid confinement, when one interface is treated with hard-wall BCs and the other with soft-wall BCs.
In Fig.~\ref{fig:hybridmodes}, we present the dispersion curve for a TI strip with hybrid BCs, where $M''=\infty$ and for $M'=10$, $100$ and $1000$~meV.
Due to the lack of the mirror symmetry the edge dispersion curves display minima at finite $k_x$.
the effect of increasing $M'$ is to partially compensate the imbalance between the two boundaries ($M''=\infty$). 
As a consequence the minima shift towards $k_x=0$ and the center of the gap tends to the hard-wall Dirac point value in Fig.~\ref{fig:Dirac}.
The gap is also increasing because of the stricter confinement from $M'$.      
In Fig.~\ref{fig:hybridspinor}, we plot the edge modes profile for the case $M'=100$~meV of Fig.~\ref{fig:hybridmodes}, corresponding to the points a, b, c  and d. 

%

\subsection{Laterally coupled TI edge states}
\begin{figure}[bph]
    \centering	
    \vspace{0.1cm}
    \includegraphics[width=6.8cm]{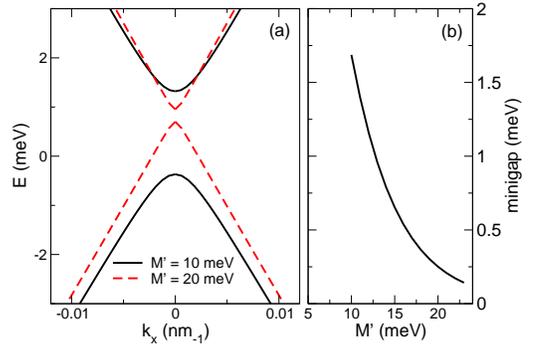}
   	\caption{(Color online) In (a), edge modes for a TI/NI/TI mass domain with width of the central NI region of $L=100$~nm and NI mass of $M'=10$ and $20$~meV are shown.
        In (b), we present the value of the minigap as a function of $M'$.
       }
    \label{fig:lateral}
 \end{figure}

We consider also the possibility to laterally couple edge modes in a TI/NI/TI mass domain, 
where the overlap of the edge modes takes place in the central NI region [see Fig.~\ref{fig:domains}(c)].
For concreteness, we assume TI regions with equal Dirac masses of $M=-10$~meV, while the NI mass is $M'>0$.
This situation leads to qualitatively similar edge modes as in a TI strip with soft-wall confinement, with the opening of a minigap at $k_x=0$ as shown in Fig~\ref{fig:lateral}(a).
However, from a quantitative point of view, edge modes decay differently in the NI region, according to their characteristic 
penetration lengths [see the inset in Fig.~\ref{fig:Dirac}].  
In Fig.~\ref{fig:lateral}(b), we display the minigap value which exponentially shrinks as a function of the the NI mass $M'$. 
%

\section{Three mass domain system}
In this section we analyze the edge states for a mass domain NI/TI/NI/TI with three boundaries [see Fig.~\ref{fig:domains}(d)] with Dirac masses
$M''>0$, $M$, $M'>0$ and $M$, respectively.
For simplicity we keep the same Dirac mass for the TI regions $M=-10$~meV.
The three-boundary system can only be realized if $M'$ is finite, while $M''$ can be either finite or $\infty$ leading to soft-wall or hard-wall BCs at the first NI/TI interface.
The edge dispersion curves are quite complex and are more easily understood by first considering the system assuming uncoupled edge states at each of the TI/NI interfaces.
Without coupling, at each TI/NI interface we expect linear edge modes similar to that in Fig.~\ref{fig:energy}, where the energy value of the Dirac point depends on the 
difference of the absolute value of the masses between NI and TI as described by Fig.~\ref{fig:Dirac}.
When edge states belonging to different TI/NI interfaces overlap a minigap is formed due to anticrossing of the dispersions.
%
The anticrossing takes place at $k_x=0$ if the two edge modes have equal Dirac point values (i.e. the two TI/NI interfaces share the same parameters), 
otherwise the anticrossing happens at finite a $k_x$.      
%
\begin{figure}[tb]
    \centering	
    \vspace{0.25cm}
    \includegraphics[width=7cm]{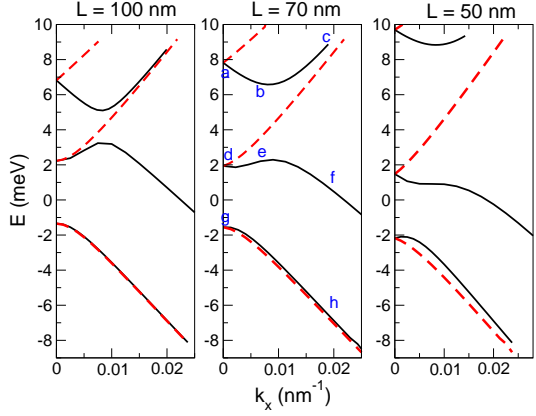}
    \caption{Edge mode dispersion curves for a NI/TI/NI/TI mass domain with three boundaries [see Fig.~\ref{fig:domains}(d)], calculated for a width of the first TI region $L=100$, $70$ and $50$~nm.
    The width of the second NI region is $d=70$~nm with Dirac mass $M'=10$ meV, while the first NI/TI interface is treated with hard-wall BCs. The second TI region is semi-infinite. Full and dashed lines distinguish $\tau=\pm 1$.
}
    \label{fig:3edges3}
\end{figure}

We focus our analysis on a system where the first NI has a very large bandgap (e.g. the vacuum) and send $M''\rightarrow\infty$ (hard-wall BCs).
We define $L$ as the width of the first TI domain (the second TI domain is considered semi-infinite) and $d$ the width of the second NI region (the one with mass $M'$).
We note that this situation is qualitatively analogous to that of a HgTe QW in a  TI phase.
The first NI/TI interface is the physical edge at the interface vacuum/HgTe QW, correctly described with hard-wall BCs.
The second NI region can be due to a large-scale (tens of nanometers in the 2D-plane) fluctuation in the QW thickness, leading to the appearance of a topologically trivial region.
HgTe QWs are typically grown to have a thickness around the critical value of $6.3$~nm and it is generally 
sufficient to have a variation of the thickness of the order of fractions of a nanometer to induce a band inversion into a NI system.

\begin{figure}[tb]
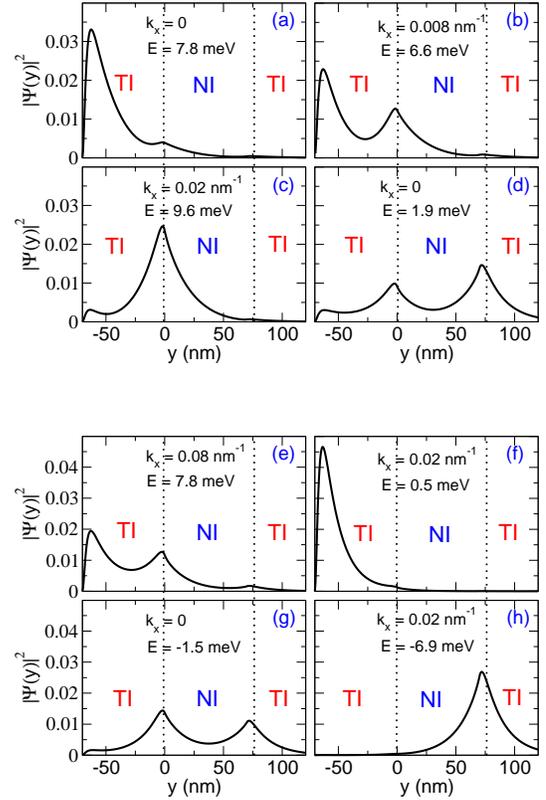

    \centering	
    \includegraphics[width=7cm]{fig/spinors2sw1hwA}\\\vspace{0.9cm}
    \includegraphics[width=7cm]{fig/spinors2sw1hwB}
    \caption{(Color online) Electronic density of edge states for a  NI/TI/NI/TI mass domain [see Fig.~\ref{fig:domains}(d)] with hard-wall BCs at the first NI/TI interface ($y=0$).
    The width of the first (from left to right) TI domain is $L=70$~nm, while that of the second NI domain is $d=70$~nm.  The Dirac mass of the second NI domain is $M'=10$ meV. The density profiles in the panels (a)-(h) correspond to states labeled by letters a-h in Fig.~\ref{fig:3edges3}.
    }
    \label{fig:3edgesspinors}
\end{figure}

In Fig.~\ref{fig:3edges3}, we plot the edge modes for a mass domain with $M'=10$~meV and $d=70$~nm and three different values of $L=100$, $70$ and $50$~nm.
The edge mode of the first NI/TI interface would have by itself a Dirac point at around $6.5$~meV, however the overlap with the edge modes from the second TI/NI interface,
which originates an anticrossing at $k_x\approx0.01$~nm$^{-1}$, pushes it up to around $8$~meV.
This analysis is confirmed by the fact that the electronic density calculated at point $a$ and $f$ of Fig.~\ref{fig:3edges3} for the case with $L=70$~nm 
[shown in Fig.~\ref{fig:3edgesspinors}(a) and (f)]  is strongly peaked near $y=0$, corresponding to the first NI/TI boundary with hard-wall BCs.
The anticrossing and the minigap opening at $k_x\approx0.01$~nm$^{-1}$ are due to the overlap of edge states between the first 
and the second boundaries as observed in the spinors in Fig.~\ref{fig:3edgesspinors}(b) and (e). 
The gap opening at $k_x=0$ around $E=0$ is due to the overlap of the edge modes of the second and third boundary and the edge states resemble the case of the lateral 
coupling of edge states through a narrow NI region with mass $M'=10$~meV (analyzed in Fig.~\ref{fig:lateral} and related text), 
as can be seen in Fig.~\ref{fig:3edgesspinors}(d) and (g).
Spinors in Fig.~\ref{fig:3edgesspinors}(c) and (h), which belong to the points $c$ and $h$ of Fig.~\ref{fig:3edges3} for the case with $L=70$~nm, 
resemble the edge modes of a symmetric TI strip with soft-wall BCs away from the minigap region of the dispersion curve.

\begin{figure}[tb]
     \centering	
     \includegraphics[width=7cm]{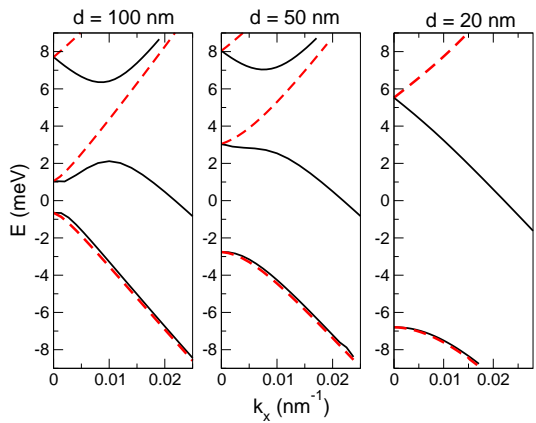}
     \caption{(Color online) Edge mode dispersion curves for a NI/TI/NI/TI mass domain with three boundaries, calculated for a width of the second NI region $d=100$, $50$ and $20$~nm 
     for a fixed Dirac mass $M'=10$ meV.
     The width of the first TI region is $L=70$~nm, while the first NI/TI interface is treated with hard-wall BCs, and the second TI region is semi-infinite. Full and dashed lines distinguish $\tau=\pm 1$.
 }        
     \label{fig:3edges2}
\end{figure}

The effect of the overlap of edge modes belonging to the second and third boundaries can be analyzed by varying $d$, which is done in Fig.~\ref{fig:3edges2}.
Decreasing $d$ from $100$~nm to $50$~nm accentuates the anticrossing behavior at $k_x=0$ around $E=0$~meV due to the overlap of edge states 
belonging to the second and the third interfaces.
Other features of the edge dispersion are only slightly affected.
When the second NI region is thinner (see case with $d=20$~nm) the overlap of the edge states bound to it is so strong that they are energetically pushed into the bulk spectral range.
As a result, the helical edge modes belonging to the first NI/TI interface (Dirac point at around 6.5 meV) are hardly affected by the presence of a second thin NI region.

\begin{figure}[tbp]
     \centering	
     \vspace{0.1cm}
     \includegraphics[width=7cm]{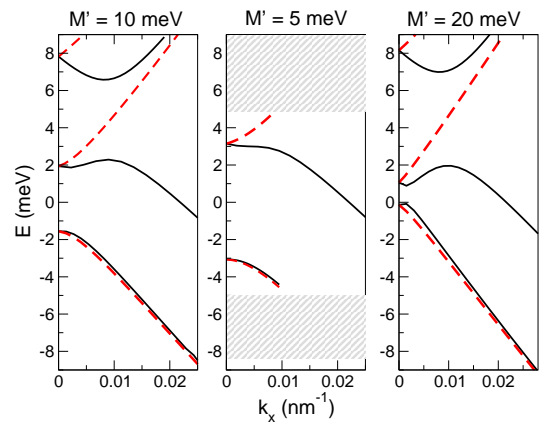}
     \caption{(Color online) Edge mode dispersion curves for a NI/TI/NI/TI mass domain with three boundaries calculated for a Dirac mass of the second NI region $M'=10$, $5$ and $20$~meV.
      The width of the first TI region is $L=70$~nm, the width of the second NI region is $d=70$~nm, while the first NI/TI interface is treated with hard-wall BCs, 
      and the second TI region is semi-infinite. Full and dashed lines distinguish $\tau=\pm 1$.
 }    
\label{fig:3edges4}
\end{figure}

The effect of varying the NI mass $M'$ is instead shown in Fig.~\ref{fig:3edges4}.
%
%
The main effect of increasing $M'$ is the reduction of the edge state coupling through the NI strip with a corresponding decrease of the $k_x=0$ minigap at $E=0$. 
If we instead decrease $M'$ the minigap at $k_x=0$ increases and the 1D edge states are restricted to a smaller spectral region since $|E|<M'$.    
%

As a final point, we note that the system with three interfaces, having an odd number of helical edge modes per spin
is protected by TRS from opening a full gap in the edge state spectrum notwithstanding the finite overlap of individual edge states.
%
%
At any given energy within the bulk-gap and for any given spin, there is an odd number (one or three) of propagating edge modes per spin.    
This is a direct consequence of the conservation of {\it the parity} of helical edge states linked to the $\mathbb Z_2$ topological invariant.
To put these statements in relation to the configurations treated in this work, we note that the configuration shown in Fig.~\ref{fig:domains}(a) with edge dispersions in Fig.~\ref{fig:energy} 
and the configuration shown in Fig.~\ref{fig:domains}(d) with edge dispersions shown in Figs.~\ref{fig:3edges3},~\ref{fig:3edges2},~\ref{fig:3edges4} have both an {\it odd} number of Dirac mass domains
(one and three, respectively) and correspondingly they are metallic. The configuration in Fig.~\ref{fig:domains}(b) with dispersions in Figs.~\ref{fig:hw} and \ref{fig:hybridmodes} and the configuration 
in Fig.~\ref{fig:domains}(c) with dispersions in Fig.~\ref{fig:lateral}(a) have an {\it even} number of Dirac mass domains and are gapped---~correspondingly they are insulators.

\section{Conclusion}
We have analyzed the edge states of a system described by the BHZ model where the Dirac mass varies spatially thus forming Dirac mass domains where topological insulating regions alternate with normal insulating regions.
While for a TI/vacuum interface the use of hard-wall boundary conditions can be assumed, we show that at a TI/NI mass domain with a finite NI mass, 
soft-wall boundary conditions (characterized by the continuity of the spinor and its derivative) are required to correctly account for the edge state dispersion curves and for the shape of the corresponding wave functions.
We solve the edge states for a system up to three TI/NI interfaces.
For the case of two interfaces, we solve the problem of a TI strip with hard-wall, soft-wall and hybrid boundary conditions, extending the work in Ref.~\onlinecite{zhou2008}.
We also have investigated the case of edge states that are laterally coupled via a narrow NI domain.  
While the edge mode spectrum is fully gapped in the two-boundary cases due to the edge mode overlap, we show that, as required by time-reversal symmetry, in the three  
boundary system, an odd number of edge modes (one or three) per spin is always present at any given energy within the bulk gap.
The models solved in this work should be relevant to understand multiple Dirac mass domains induced by fractions of nanometer ranged thickness fluctuations in HgTe-based 
quantum wells or via tunable voltage-induced band-inversions, e.g. in double quantum well structures~\cite{michetti2012}.  
Such a controlled creation of multiple helical edge states within a single structure could be used to create tunable spin- and charge-transport devices \cite{Liu2011, Krueckl2011, Dolcini2011, Romeo2012}.

\begin{acknowledgments}
We acknowledge useful discussions with Grigory Tkachov and financial support from the DFG grant RE 2978/1-1 (P.M. and P.R.) and from CNPq, CAPES and 
FAPESP (P.H.P and J.C.E). P.H.P. acknowledges the kind hospitality at the Institute of Theoretical Physics and Astrophysics, University of W\"urzburg, 
where part of this work was developed, and financial support for her visit.
\end{acknowledgments}

  \appendix
  \section{The 4-band Model~\label{app1}}
  We consider the usual TI Hamiltonian for one of the Kramers partners, given by the following 
  \begin{equation}
    h({\vec{k}})\hspace{-0.05cm}  = \hspace{-0.03cm} \left( C \hspace{-0.07cm} -\hspace{-0.07cm} D k^2\right)\hspace{-0.05cm} \sigma_0 \hspace{-0.03cm} 
                                  + \hspace{-0.03cm} ( M\hspace{-0.06cm} -\hspace{-0.06cm} B k^2)\sigma_z   \hspace{-0.03cm} 
                                  + \hspace{-0.03cm}  A (k_x\sigma_x\hspace{-0.03cm}-\hspace{-0.03cm}k_y\sigma_y).
  \end{equation}
  Let us define the following two conditions 
  \begin{eqnarray}
    \nu M < \nu \frac{A^2}{4B},  
    \label{condition1}\\
    (k_{min}^{\pm})^2>0,
    \label{condition2}
  \end{eqnarray}
  where $\nu={\rm sgn}\left[B/(D^2-B^2)\right]$. 
  If both conditions in Eqs.~(\ref{condition1}) and (\ref{condition2}) are satisfied, the bulk energy dispersion, 
  given by Eq.~(\ref{eq:bulk}), has a ``Mexican hat'' form with a local maximum at $k=0$, and
  \begin{equation}
    (k_{min}^{\pm})^2 = \frac{M}{B} - \frac{A^2}{2B^2} \pm \frac{|A| D}{2B^2} \sqrt{\frac{A^2 -4 M B}{D^2-B^2}},
  \end{equation}
  are the valence band maxima ($-$) and conduction band minima ($+$), respectively, with energies
  \begin{equation}
    \varepsilon_{\pm}^{C}\hspace{-0.03cm}  =\hspace{-0.03cm}  -\frac{D\hspace{-0.01cm}  M}{B} \hspace{-0.01cm} + \hspace{-0.01cm}  \frac{DA^2}{2\hspace{-0.01cm} B^2} \hspace{-0.01cm} \pm\hspace{-0.01cm}  
    \frac{|A|}{2\hspace{-0.01cm}  B^2} \sqrt{\hspace{-0.05cm}  (\hspace{-0.02cm} A^2 \hspace{-0.05cm} -\hspace{-0.05cm} 4\hspace{-0.01cm}  M\hspace{-0.01cm}  B\hspace{-0.01cm} )(\hspace{-0.01cm} D^2\hspace{-0.05cm} -\hspace{-0.05cm} B^2\hspace{-0.01cm} )}.
    \label{minimi}
  \end{equation}

  If Eq.~(\ref{condition1}) is satisfied but Eq.~(\ref{condition2}) is not for $k_{min}^{+}$($k_{min}^{-}$) then still the conduction (valence) band has a single minimum at $k=0$ with energy $|M|$ ($-|M|$).  

  With standard TI QW parameters the condition in Eq.~(\ref{condition1}) is generally not fulfilled; with our choice of parameters it would correspond to
  \begin{equation}
    M < \frac{A^2}{4 B}\approx - 32~{\rm meV},
  \end{equation}
  and therefore the valence (conduction) band has a maximum (minimum) at $k=0$ and energy $E=M$ ($E=-M$).
 
  Compatible with a fixed energy $E$ and a real $k_x$, one generally obtains four complex values of $k_y$, given by Eq.~(\ref{modi}).
  Let us now analyze the domain of $k_y$ as a function of the TI parameters.
  Eq.~(\ref{modi}) leads to complex solutions if
  \begin{eqnarray}
    F^2-Q^2 
    <0,
    \label{condition3}
  \end{eqnarray}
  otherwise the solutions are either purely real or purely imaginary.
  %
  %
  %
  Such an analysis shows that complex $k_y$'s are found for $E \in \left( \varepsilon_-^{C},\varepsilon_+^{C}\right)$ if $|B|>|D|$ and  
  in $E \notin\left( \varepsilon_-^{C},\varepsilon_+^{C}\right)$ for $|B|<|D|$ . 
 
  Note that for $|B|>|D|$, if both conduction and valence band have a Mexican hat form, then complex $k_y$'s are essentially found inside the gap region, 
  bound by $\varepsilon^{C}_{\pm}$ in Eq.~(\ref{minimi}).
  In this case, $k_y=k_\pm$ [defined in Eq.~(\ref{modi})] are both complex in the gap spectral range with $|k_+|=|k_-|$.
  We define
  \begin{eqnarray}
    \tilde{k}_\pm\hspace{-0.05cm}  \doteq  \hspace{-0.02cm} \pm \hspace{-0.01cm} k_\pm\hspace{-0.02cm}  
                                   =       \hspace{-0.02cm} \pm \hspace{-0.01cm} \sqrt{\hspace{-0.03cm} -k_x^2 \hspace{-0.05cm} - \hspace{-0.03cm} F \hspace{-0.02cm} 
                                          \pm \hspace{-0.03cm}  i \hspace{-0.01cm}  \sqrt{\hspace{-0.02cm} |\hspace{-0.01cm}F^2\hspace{-0.05cm} -\hspace{-0.05cm} Q^2|}}\hspace{-0.05cm} 
                                    =    \hspace{-0.01cm} \pm u \hspace{-0.01cm} +\hspace{-0.01cm}  i v,
    \label{complex}
  \end{eqnarray}
  where $u$ and $v$ are the real and imaginary parts of $k_{\pm}$ and $v>0$.
  We choose the present definition of $\tilde{k}_\pm$, so that they exponentially decay along the $y$-axis.
  Components corresponding to $k_y=\tilde{k}_\pm$ contribute to edge states (if existent) with a single decay length $1/v$ and an oscillatory behavior as $\sin{(uy)}$.
  If Eq.~(\ref{condition1}) is satisfied but Eq.~(\ref{condition2}) is not for $k_{min}^{+}$($k_{min}^{-}$) then in the 
  interval between $\varepsilon_{+}^C$ ($\varepsilon_-^C$) and $-|M|$ ($|M|$) $k_y$ is purely imaginary.
  In this case we, instead, define $\tilde{k}_\pm \doteq k_\pm$ and, when existent, edge states will have two decaying 
  lengths ($1/k_\pm$) and no oscillatory behavior~\cite{Lu2012, michetti2012}.


\begin{thebibliography}{24}
\bibitem{footnote2} 
To make the notation more compact, we use four component Pauli matrices  $\vec{\sigma}=\left(\mathbb{I}, \sigma_x, \sigma_y, \sigma_z \right)$.
\bibitem{adiabatic} 
If Dirac mass domains are induced by thickness variations of QWs, the width over which the mass domains develop should be smooth for the out-of-plane ($z$-axis) 
dynamics in order to avoid mixing with higher subbands, but still step-like for the in-plane motion which is governed by Eq.~(1).
Note that the typical in-plane extent of the edge states is of the order of $40$~nm, 
therefore we assume the Dirac mass domain to develop on a much shorter length scale.
\end{thebibliography}

\end{document}